\documentclass[aps,prl,twocolumn,preprintnumbers,showpacs,nofootinbib]{revtex4}

\usepackage[spanish]{babel}
\usepackage{amsmath}
\usepackage{amssymb}
\usepackage{epsfig}
\usepackage{graphics,psfrag,rotating}
\usepackage{graphicx}% Include figure files
\usepackage{dcolumn}% Align table columns on decimal point
\usepackage{bm}% bold math
\begin{document}
\newcommand{\Od}{{\cal O}}
\newcommand{\lsim}   {\mathrel{\mathop{\kern 0pt \rlap
  {\raise.2ex\hbox{$<$}}}
  \lower.9ex\hbox{\kern-.190em $\sim$}}}
\newcommand{\gsim}   {\mathrel{\mathop{\kern 0pt \rlap
  {\raise.2ex\hbox{$>$}}}
  \lower.9ex\hbox{\kern-.190em $\sim$}}}

%\preprint{APS/123-QED}

\title{A $f(R)$ gravity without cosmological constant}%
\author{\'Alvaro de la Cruz-Dombriz\footnote{E-mail: dombriz@fis.ucm.es}
and Antonio Dobado\footnote{E-mail: dobado@fis.ucm.es}}

\affiliation{Departamento de  F\'{\i}sica Te\'orica I,
 Universidad Complutense de
  Madrid, 28040 Madrid, Spain}%

\date{\today}% It is always \today, today,


\begin{abstract}
In this work we consider the possibility of describing the current
evolution of the universe, without the introduction of any
cosmological constant or dark energy (DE), by modifying the
Einstein-Hilbert (EH) action. In the context of the $f(R)$
gravities within the metric formalism, we show that it is possible
to
 find an action without cosmological constant which exactly
reproduces the behavior of the EH action with cosmological
constant. In addition the $f(R)$  action is analytical at the
origin having Minkowski and Schwarzschild as vacuum solutions. The
found $f(R)$ action is highly non-trivial and must be written in
terms of hypergeometric functions but, in spite of looking
somewhat artificial, it shows that the cosmological constant, or
more generally the DE, is not a logical necessity.
\end{abstract}

\pacs{98.80.-k, 04.50.+h}% PACS, the Physics and Astronomy
\maketitle

One of the most important recent scientific discoveries is the
accelerated expansion of the universe. Different data from
type Ia supernovae \cite{typeIa1, typeIa2, typeIa3} observation,
large structure information and delicate measurements of the
cosmic microwave background (CMB) anisotropies (particularly those
from the Wilkinson Microwave Anisotropy Probe (WMAP)  \cite{WMAP})
have concluded
 that our universe is expanding at an
increasing rate. This fact sets the very urgent problem of finding
the cause for this speed-up.

Usual explanations belong to one of the following three classes:
First one reconciles this acceleration with General Relativity
(GR) by invoking a strange cosmic fluid, DE, with a state equation
$ p =\omega \rho$ where $\omega$ is very close to -1, i.e. the
fluid has a large negative pressure. For the particular case
$\omega=-1$ this fluid behaves just as a cosmological constant
$\Lambda$. Within this approach recent data obtained by WMAP correspond
to the cosmological parameters \cite{WMAP}:
$\Omega_{M}h^2=0.1493^{+0.007}_{-0.013}$, $\Omega_{\Lambda}
=0.72\pm0.04$ and $H_{0}(t=t_{0})=100 h\, km \,s^{-1} Mpc^{-1}$
with
  $h=0.71^{+0.03}_{-0.03}$ and $t_{0}\equiv t_{today}$. The main
  problem of this kind of description is that the fitted $\Lambda$  value
   seems to be about 55 orders of magnitude smaller than
   expected (the cosmological constant problem). The second type of
   explanations consider a dynamical DE by introducing
    a new scalar field. Finally the third one is trying to
    explain the cosmic acceleration as a consequence of new
    gravitational physics \cite{Carroll1}. EH action modifications have been widely considered in the
    literature \cite{S,B,M}, firstly to describe inflation, and more recently to
    describe the current cosmic speed-up, or even both cosmological eras simultaneously.

    The simplest
    way of modifying EH action is by adding
    some function $f(R)$ with the required properties ($f(R)$ gravities).
    For example in \cite{Carroll2} it was introduced a gravitational model
    where $f(R)=-\mu^4/R$, being the total gravitational action proportional to
    $R-\mu^4/R$.  This proposal has very interesting cosmological properties and
     triggered a lot of work on $f(R)$ gravities applied to
     cosmology. However this kind of actions with negative powers of the
     curvature has the very serious drawback of not having vacuum
     solutions with vanishing curvature. For instance in the mentioned model the vacuum constant curvature solution is
     $R=\pm  \sqrt{3}\mu^2$. Thus, even if one succeeds in reproducing cosmic acceleration,
      paradigmatic GR vacuum
     solutions assumed to play a major role in any fundamental theory of
     gravity, such as Minkowski or Schwarzschild, are excluded. Other $f(R)$
      functions recently considered in the literature face similar problems and moreover
      could be in conflict with Solar System experiments
      \cite{Olmo1,Cembranos}.

     In this work we address the issue of finding a $f(R)$ gravity
     able to reproduce the current cosmic speed-up without any cosmological constant
     but having $R=0$ as vacuum solution. From a more formal
     point of view we are seeking for a $f(R)$ gravity having the same Friedmann-Robertson-Walker (FRW)
     solution as the standard EH action with cosmological constant
     for non relativistic matter ($p=0$), but being analytical at
     $R=0$. Clearly the $f(R)$ expansion at $R=0$ must start at the $R^2$ term
     to avoid having cosmological constant or to redefine the Newton
     constant.

In order to consider such as the standard EH cosmological solution
with DE as $f(R)$ gravity cosmological solution without DE in the
same setting, we start from a general action $S=S_G+S_M+S_{DE}$
where $S_G$ is the gravitational action given by:
\begin{equation}
    S_G=\frac{1}{2\kappa}\int d^4x \sqrt{\vert g \vert}[R+f(R)]
\label{Action}
\end{equation}
with $\kappa\equiv8\pi G_N$ and $f(R)$ being any arbitrary function
of the scalar curvature $R\,=\,g^{\mu\nu}R_{\mu\nu}$, where the
Ricci tensor is given by $R_{\mu\nu}\,=\,R_{\mu\alpha\nu}^{\alpha}$
and the curvature tensor is defined with the convention $
R_{\beta\mu\nu}^{\alpha}\, \sim\,\partial_{\beta}
\Gamma^{\alpha}_{\mu\nu}$ where $\Gamma^{\alpha}_{\mu\nu}$ are the
symbols for the Levi-Civita connection since we are using the metric
formalism (see \cite{Sotiriou} for a recent review on non-metric
formalisms). In addition, $S_M$ is the action describing matter
(also including dark matter) and $S_{DE}$ is the action describing 
DE, which in particular includes any possible cosmological constant
$\Lambda$. For these two actions the corresponding energy-momentum
tensor are given by $T_X^{\mu\nu}= 2\vert g \vert^{-1/2}\delta
S_{X}/\delta g_{\mu\nu}$ with $X=$ M (matter) or $X=$ DE. For the
sake of simplicity DE will be assumed to follow
$p_{DE}\,=\,-\rho_{DE}$ as state equation (i.e. it is just a
cosmological constant) where $\rho_{DE}\equiv\Lambda/\kappa$. Thus
in our notation
$(T_{DE})^{\mu}_{\,\,\,\nu}=-\rho_{DE}\delta^{\mu}_{\,\,\,\nu}$.
Assuming that
 matter (including dark matter) can be
described as a perfect fluid, the corresponding energy-momentum
tensor is $(T_{M})^{\mu}_{\,\,\,\nu}=-\,diag(\rho_M , -p_M , -p_M
, -p_M)$.

In the metric formalism field tensorial equations are found by
performing variations of the above action \eqref{Action} with
respect
 to the metric. Thus the equations are:
\begin{equation}
[1+f'(R)]R_{\mu\nu}-\frac{1}{2}[R+f(R)]g_{\mu\nu}+{\cal
D}_{\mu\nu}f'(R)\,=\,\kappa T_{\mu\nu}
\label{fieldtensorialequation}
\end{equation}
where $'$ represents the derivative with respect to $R$, ${\cal
D}_{\mu\nu}\equiv D_{\mu}D_{\nu}-g_{\mu\nu}\square$,
$\square\,\equiv\,D_{\alpha}D^{\alpha}$ and $D$ is the usual
covariant derivative. By computing covariant derivative of
\eqref{fieldtensorialequation} we find the equations of motion
$D_{\mu}T^{\mu}_{\,\,\,\nu}=0$ independently from $f(R)$ and
$\Lambda$.
 Clearly for the particular case $f(R)\equiv0$ we recover the standard Einstein field equations
with cosmological constant, i.e.
\begin{eqnarray}
R_{\mu\nu}-\frac{1}{2}g_{\mu\nu}R+ g_{\mu\nu}\Lambda=\,\kappa
(T_{M})_{\mu\nu} \label{einsteinfieldequation}
\end{eqnarray}
and the matter equation of motion
$D_{\mu}(T_{M})^{\mu}_{\,\,\,\nu}=0$. In the following we will be
interested in the cosmological solutions of the above equations
with flat spatial sections. Thus we will consider the line element
\begin{equation}
ds^2\,=\,dt^2-a^2(t)(dr^2+r^2d\Omega_{2}^2) \label{metric}
\end{equation}
From this metric the matter equation of motion in the general case
reads:
\begin{equation}
\dot{\rho}_M+3(1+\omega_M)\rho_M\frac{\dot{a}}{a}\,=\,0
\label{eq_dif_densidad}
\end{equation}
with $\partial_{k}\rho_M=0$ where $k$ runs through $r$, $\theta$
and $\phi$, the dot represents the time derivative and we have
assumed the matter state equation to be $p_M=\omega_M\rho_M$.
These two equations can be integrated to give:
\begin{equation}
\rho_M(t)\,=\,\rho_M(t_{0})\left(\frac{a(t_{0})}{a(t)}\right)^{3(1+\omega_M)}
\label{matterdensity}
\end{equation}
where $t_0$ is the present time. Now from the $\mu,\nu=t$
component of \eqref{fieldtensorialequation} we have:
\begin{equation}
R^{t}_{\;\;t}[1+f'(R)]-\frac{1}{2}[R+f(R)] +\Lambda +{\cal
D}^{t}_{\,\,\,t}f'(R)\, = \,  - \kappa\rho_M.
 \label{00}
\end{equation}
In a flat universe it is possible to write the scalar curvature in
terms of scale parameter $a\equiv a(t)$ and its derivatives as:
\begin{equation}
R\,=\,6\left[\left(\frac{\dot{a}}{a}\right)^2 +
\frac{\mbox{\"a}}{a}\right] \label{R}
\end{equation}
and subsequently, for those universes
 $R^{t}_{\;\; t}=3\mbox{\"a}/a$ and ${\cal D}^{t}_{\,\,\,t}f'(R)\,=\,-3\dot{a}\dot{R}f''(R)/a$.
The most recent cosmological data quoted in the introduction are
compatible with a cosmological model based on a flat FRW metric like
\eqref{metric} together with Einstein field equations with
cosmological constant $\Lambda\neq0$ and non relativistic (dust)
matter (including dark matter), i.e. $p_M=0$. In this case we can
use the equation \eqref{fieldtensorialequation} with $f(R)=0$ and
$T^{\mu}_{\,\,\,\nu}=-\,diag(\rho_M +\rho_{DE}, -\rho_{DE} ,
-\rho_{DE} , -\rho_{DE})$. Then we find:
\begin{equation}
\left(\frac{\dot{a_{0}}}{a_{0}}\right)^2\,=\,\frac{\kappa}{3}\rho_{M0}+\frac{\Lambda}{3}
\label{00con}
\end{equation}
where the $0$ subindex means that we are using standard EH cosmology
equations with cosmological constant. This notation will be relevant
later on when we will compare standard cosmology with the results
coming from the $f(R)$ action for gravity that we have found in this
work. The above equation can be solved exactly to find:
\begin{eqnarray}
a_{0}(t)\,&=&\,\left(\frac{\Omega_{M}}{\Omega_{\Lambda}}\right)^{1/3}sinh^{2/3}
\left[\frac{3}{2}H_{0}(t_{0})(t-t_{0}) \sqrt{\Omega_{\Lambda}}+
\right. \nonumber\\ &+& \left.
arcsinh[\sqrt{\frac{\Omega_{\Lambda}}{\Omega_{M}}}]\right]
\label{00_HE_resuelta}
\end{eqnarray}
where we have used the notation:
 $H_{0}(t)\equiv \dot{a_0}(t)/a_{0}(t)$,
 $\Omega_{\Lambda}\equiv \Lambda/3H_{0}^2 (t_{0})$ and
  $\Omega_{M}\equiv \kappa\rho_{M0}(t_{0})/3H_{0}^2(t_0)$ with the
   condition $a_0(t_0)\,=\,1$. On the other hand, by taking
   the trace of \eqref{fieldtensorialequation} in this case, ie.
   $f(R)\equiv0$ and dust, we find:
\begin{equation}
R_{0}(t)-4\Lambda\,=\,\kappa\rho_{M0}(t) \label{traza_con_f_nula}
\end{equation}
Now we consider again equation \eqref{fieldtensorialequation} but
in the case where we have an arbitrary function $f(R)$ in the
action but not DE, i.e. we set $\Lambda\equiv0$  so
 that $T^{\mu}_{\,\,\,\nu}=-\,diag(\rho_M ,0,0,0)$.
Then we have:
\begin{equation}
3[1+f'(R)]\frac{\mbox{\"a}}{a}-\frac{1}{2}
[R+f(R)]-3\frac{\dot{a}}{a}\dot{R}f''(R)\,= \,-\kappa\rho_{M}
\label{00sin}
\end{equation}
where we have eliminated the subindex $0$ in the different
quantities to avoid any confusion with the previous case. At this
stage it is clear
 that above equation \eqref{00sin} solutions
will depend on the function $f(R)$: different choices for this
function will lead to different evolutions of the universe for the
same initial conditions. However, our approach to the problem will
be to find a function $f(R)$ so that the solution $a(t)$ of
\eqref{00sin} will be exactly the same as the solution in
\eqref{00_HE_resuelta} that we got by using standard cosmology and
which fits the present cosmological data. In other words we want
to find a $f(R)$ such that:
\begin{equation}
a(t)=a_0(t)
\end{equation}
for the same initial (or better present, i.e. $t=t_0$,
conditions). If it were possible to find this function $f(R)$ then
it would be possible to avoid the necessity for introducing any
cosmological constant just by modifying
 the gravitational sector of the action. In the following we will show that such a
function happens to exist and we will give its precise form. In
order to do that we first notice that having $a(t)\equiv a_0(t)$
in this period of universe life clearly implies $R(t) \equiv
R_{0}(t)$ and then we can substitute $R$ by $R_0$ in
\eqref{00sin}. On the other hand we will write the matter density
as the former matter density plus a new
 contribution, ie. $\rho_{M}(t)\,=\,\rho_{M0}(t) +\Delta\rho(t)$.

Assuming that matter for arbitrary $f(R)$ is still
non-relativistic (i.e. dust) in this cosmological era we have
\begin{equation}
\Delta\rho(t)\,=\,\Delta\rho(t_{0})\left(\frac{a_{0}(t_{0})}{a_{0}(t)}\right)^3
\label{delta_densidad_escala}
\end{equation}
where according to \eqref{matterdensity} particularized for
$a\,=\,a_0$, and \eqref{traza_con_f_nula} we can write
\begin{equation}
\left(\frac{a_{0}(t_{0})}{a_{0}(t)}\right)^3=\frac{R_{0}-4\Lambda}{\kappa\rho_{M0}(t_{0})}
\label{factor_de_escala}
\end{equation}
and then \eqref{delta_densidad_escala} becomes
\begin{equation}
\Delta\rho(t)=-\eta\frac{R_{0}-4\Lambda}{\kappa}
\label{delta_rho_final}
\end{equation}
where we have introduced the parameter $ \eta \equiv
-\Delta\rho(t_{0})/\rho_{M0}(t_{0})$ so that matter density is
written as $\rho_{M}(t;\eta)=(1-\eta)\rho_{M0}(t)$. Finally the
last term on the l.h.s. of
 \eqref{00sin} can be written in terms of the scalar curvature
by differentiating  \eqref{traza_con_f_nula} and using
\eqref{eq_dif_densidad}. Thus we get
\begin{eqnarray}
&3&\left[R_0-3\Lambda\right]\left[R_0-4\Lambda\right]f''(R_0)  +
 \left[-\frac{1}{2}R_0+3\Lambda\right]f'(R_0) - \nonumber\\
&-&\frac{1}{2}f(R_0)-\Lambda - \eta \left[R_0-4\Lambda\right]
\,=\,0 \label{e_omega=0}
\end{eqnarray}
This last equation can be considered as a differential equation for
the function $f(R)$ (in the following we will omit the subindex $0$
in $R$ since no confusion is possible). \eqref{e_omega=0} is a
second order linear equation so two initial conditions are needed to
solve it: $f(0)$ and $f'(0)$ for instance. The natural choice for
these initial conditions will be the following: Firstly we do not
want to have any cosmological constant in our action, so that
$f(0)=0$. Secondly we want to recover the standard EH action for low
scalar curvatures without redefine the Newton constant, i.e.
 $f'(0)=0$. Moreover we want $f(R)$ to
be an analytical function at the origin so that $R=0$ should be a
solution for the field equations in vacuum. This is an extremely
important requirement since it allows Minkowski and Schwarzschild
to be vacuum solutions.

With these initial conditions \eqref{e_omega=0} can be solved by
using standard methods. A particular solution is:
\begin{equation}
f_p(R)\,=\,-\eta R+2\Lambda\left[\eta-1\right]
\label{sol_particular_completa}
\end{equation}
The homogeneous equation associated with \eqref{e_omega=0} is a
Gauss equation solved in terms of hypergeometric functions. The
general solution of the homogeneous equation can be written as:
\begin{equation}
f_h(R)\,= \Lambda[K_{+} f_{+}(R)+ K_{-} f_{-}(R)]
\end{equation}
where
\begin{equation}
f_{\pm}(R)  \,=\,  \alpha^{-a_{\pm}}
\,_{2}F_{1}[a_{\pm},1+a_{\pm}-c;1+a_{\pm}-a_{\mp};-\alpha^{-1}]
\label{sol_homogenea}
\end{equation}
with $\alpha= 3-R/\Lambda$ and
\begin{equation}
\begin{array}{l}
    a_{\pm}\,=\,-\frac{1}{12}(7 \pm \sqrt{73}), \,\,\,\,\,\,c\,=\,-1/2
\end{array}
\label{ac}
\end{equation}
The $\eta$ dependent constants $K_{+}$ and $K_{-}$ must be
determined from the initial conditions given above. Numerically we
find: $K_{+}\,=\,0.6436 ( -0.9058\eta+0.0596)$ and $
K_{-}\,=\,0.6436(-0.2423\eta+3.4465)$.

The hypergeometric functions given in \eqref{sol_homogenea} are
generally defined in the whole complex plane. However we want to
have a real gravitational action. In principle this is very easy
to achieve  since the coefficients in the equation
\eqref{e_omega=0} and the  constants $K_{+}$ and $K_{-}$ are all
of them real. Then it is obvious that the real part of
\eqref{sol_homogenea} is a proper solution of homogeneous equation
associated with \eqref{e_omega=0}. Thus the function we are
seeking can be written as:

\begin{equation}
f(R)\,\equiv\, f_p(R)+Re\left[f_h(R)\right] \label{sol_definition}
\end{equation}
Nevertheless situation is a bit more complicated. The homogeneous
equation has three regular singular points at $R_{1}\,=\,3\Lambda$,
$R_{2}\,=\,4\Lambda$ and $R_{3}\,=\,\infty$. This results in the
solution $f_h(R)$ having two branch points $R_1$ and $R_2$. More
concretely there are two cuts along the real axis: one from minus
infinity to $R_1$ and another from $R_2$ to infinity. Thus one must
be quite careful when interpreting \eqref{sol_definition}. From
minus infinity to $R_1$ there is only one Riemann sheet of $f_h(R)$
where $f(0)$ and $f'(0)$ vanish and therefore this is the one that
we have to use to define  $f(R)$. From $R_1$ to $R_2$ the real part
of $f_h(R)$ is well defined. Finally from $R_2$ to infinity there is
only one Riemann sheet producing a smooth behavior of $f(R)$. To
reach this sheet one must understand $R$ in the above equation as
$R+i\epsilon$.

At the
 present moment we do not know if this analytical structure
 has any fundamental meaning or it is just an artefact of our
construction. Much more important is the fact that the function
$R+f_p(R)+f_h(R)$, which is the analytical extension of our
lagrangian, is analytical at $R=0$, having at this point the local
behavior $R+O(R^2)$. Therefore our generalized gravitational lagrangian
$R+f(R)$ does guarantee that $R=0$ is a vacuum solution whilst it reproduces the current evolution of the
universe without any cosmological constant. Once we have obtained
the $f(R)$ function it is possible to check out our result by
solving \eqref{00sin} in terms of $a(t)$ for the $f(R)$ given in
\eqref{sol_definition}. This is done by rewriting equation
\eqref{00sin} in terms of $a(t)$ by using \eqref{matterdensity}
 and \eqref{R} together with the $\rho_{M}$ definition in terms of $\eta$. This process requires three
 initial conditions given for the present time $t_{0}$ which are: $a(t_0)=1, \dot{a}(t_0)=H_{0}(t_0)
  a(t_0)$ and $\mbox{\"a}(t_0)= -q_0 a(t_0) H_{0}^2(t_{0})$ with
$q_{0}\simeq-0.61$. The numerical solution $a(t)$ shows a
nice agreement with $a_0(t)$ given in \eqref{00_HE_resuelta}.
Thus we can be sure that our gravitational action proportional to
$R+f(R)$ provides the same cosmic evolution as EH action
$R-2\Lambda$ in a dust matter universe. Therefore, our model will
verify, in the same range of precision, all the experimental tests that the standard cosmological model verifies in the present era.

   Notice also that in principle this can achieved for any value of $\eta$, i.e. for any desired amount of
   matter. Nevertheless some restrictions should be imposed over the
    parameter $\eta$. For instance it is obvious that in a
    dust matter dominated universe  $\rho_{M}(t; \eta) \geq 0$
    implies $\eta\leq1$.

Much more stringent bounds can be set on $\eta$ by demanding our
model to work properly back in time up to Big Bang
Nucleosynthesis (BBN) era. Observations indicate that the
cosmological standard model fits correctly primordial
 light elements abundances during BBN with a 10$\%$ of relative error for $H_{0}(t)$. Therefore
  by the time of BBN, departure of our model from the standard cosmology must not be too large and
   \eqref{00sin} should  give
   similar behavior to the one given by the standard Friedmann
   equation \eqref{00con} where now $\rho_{M0}\,=\,\rho_{0}^{dust}(t)+\rho_{0}^{radiation}(t)$.

 At BBN era cosmological constant is negligible compared
 with dust and radiation densities. The scalar curvature
 is of order $10^{-39}$ $eV^2$ (with $\hbar\,=\,c\,=\,1$ for these calculations) and by that time
  dust and radiation densities are of the order of  $10^{16}$ $eV^{4}$
  and $10^{21}$ $eV^{4}$  respectively.
Since $R \simeq R_{0}$ we can rewrite \eqref{00sin} as a modified
Friedmann equation as follows
\begin{equation}
H^2(t)\,=\,H_{0}^2(t)\left[ \frac{10^{5}R-\eta
R+\frac{1}{2}[Rf'(R)-f(R)]}{10^{5}R[1+f'(R)-3f''(R)(1-\eta)R]}
\right]. \label{00sin_hubble_final}
\end{equation}
As it was commented above, to reproduce light elements abundances
it is required that $H^2(t)\,=\,H_{0}^2(t)(1\pm0.2)$ for
curvatures of order $R_{BBN}$. This implies that the second factor
on the r.h.s. of \eqref{00sin_hubble_final}  should be between
$0.8$ and $1.2$ by that period. Thus in order to match our $f(R)$
gravity model with the standard cosmology at the BBN times we need
to tune  $\eta$ to a value about $0.065$. Therefore the matter
content of our model is not too different from the one in the
standard cosmology and the difference is in fact smaller than
experimental precision in \cite{WMAP}.

To conclude we have succeeded  in finding a $f(R)$ gravity which
exactly reproduces the same evolution of the universe, from BNN up
to the present time, as standard cosmology, but without the
introduction of any form of DE or cosmological constant. The
gravitational lagrangian $R+f(R)$ is analytical at the origin and
consequently $R=0$ is a vacuum solution of the field equations.
Therefore Minkowski, Schwarzschild and other important $R=0$ GR
solutions, with $\Lambda=0$, are also solutions for this $f(R)$
gravity. The price we have to pay for all those good properties is
that our lagrangian, considered as a function of $R$, has a very
complicated analytical structure with cuts along the real axis from
infinity to $R=3\Lambda$ and from $R=4\Lambda$ to infinity.
Obviously the only reasonable interpretation of our action is as
some kind of effective action. In classical physics one typically
starts from some action principle, obtains the corresponding field
equations and finally solves them for some initial or boundary
conditions. In this work we have proceeded in the opposite way: we
started from solutions obtained in the standard cosmological model
and then we have searched for an action that, possessing certain
properties, gives rise to field equations having the same solutions.
Classical actions are of course real but effective quantum actions
usually have a complex structure coming from loops and related to
unitarity. The presence of an imaginary part in the action,
evaluated on some classical configuration, indicates quantum lost of
stability by particle emission of this configuration \cite{MD}.
Therefore it is tempting to think that our action could have some
interpretation in terms of an effective quantum action. However, our
action determination procedure does not allow to make such a kind of
statement. Complications in the action could be just an artifact of
our construction. In any case we have shown that such action exists
and it reproduces the present universe evolution without DE  having
$R=0$ as a vacuum solution. The complicated structure of this action
may be an indication that the cosmological constant problem is even
much harder to solve than we have previously thought. Obviously much
more insight and research are needed in order to get further progress in
this issue.

%\vspace{.1cm}

 {\bf Acknowledgements:} We would like to thank J.A.R. Cembranos
 and very specially A.L. Maroto for useful comments and discussions. This work
 has been partially supported by the DGICYT (Spain) under the
 project FPA2005-02327.


\begin{thebibliography}{99}

\bibitem{typeIa1} A. G. Riess {\it et al}.[Supernova Search Team
Collaboration], Astrom. J. {\bf 116}, 1009 (1998)
[arXiv:astro-ph/9805201].

\bibitem{typeIa2} S. Perlmutter {\it et al}.[Supernova Cosmology
Project Collaboration], Astrophys. J. {\bf 517}, 565 (1999)
[arXiv:astro-ph/9812133].

\bibitem{typeIa3} J. L. Tonry {\it et al}. Astrophys. J. {\bf 594}, 1 (2003)
[arXiv:astro-ph/0305008].

\bibitem{WMAP} D.N.Spergel et al. \emph{Wilkinson
Microwave Anisotropy Probe (WMAP) Three Year Results:
 Implications for Cosmology}. astro-ph/0603449, 2006.

\bibitem{Carroll1}  S.~M.~Carroll, A.~De Felice, V.~Duvvuri, D.~A.~Easson, M.~Trodden and M.~S.~Turner,
  Phys.\ Rev.\ D {\bf 71} (2005) 063513
  [arXiv:astro-ph/0410031].

\bibitem{S} %\cite{Carroll:2004de}
  A. A. Starobinsky, Phys.\ Lett.\ {\bf 91B}, 99 (1980).

\bibitem{B} J. D. Barrow and A. C. Ottewill, J.\ Phys.\ A {\bf 16},
2757 (1983).

\bibitem{M} A. Dobado and A. L. Maroto
  Phys.\ Rev.\ D {\bf 52}, 1895 (1995).

\bibitem{Carroll2}
  S.~M.~Carroll, V.~Duvvuri, M.~Trodden and M.~S.~Turner,
  Phys.\ Rev.\ D {\bf 70} (20.04) 043528
  [arXiv:astro-ph/0306438].

\bibitem{Olmo1}
  G.~J.~Olmo,
  %``The gravity lagrangian according to solar system experiments,''
  Phys.\ Rev.\ Lett.\  {\bf 95} (2005) 261102
  [arXiv:gr-qc/0505101].

\bibitem{Cembranos}
  J.~A.~R.~Cembranos,
  %``The newtonian limit at intermediate energies,''
  Phys.\ Rev.\ D {\bf 73} (2006) 064029
  [arXiv:gr-qc/0507039].

\bibitem{Sotiriou} T. P. Sotiriou and S. Liberati [arXiv:gr-qc/0604006].

\bibitem{MD}   A. Dobado and A. L. Maroto, Phys.Rev.{\bf D60}, (1999)
104045

\end{thebibliography}
\end{document}